\documentclass[12pt]{article}
\newlength{\pubnumber} 

\catcode`\@=11
\@addtoreset{equation}{section}

\def\section{\@startsection{section}{1}{\z@}{3.5ex plus 1ex minus .2ex}
 {2.3ex plus .2ex}{\large\bf}}
\def\subsection{\@startsection{subsection}{2}{\z@}{2.3ex plus .2ex}
 {2.3ex plus .2ex}{\bf}}

\begin{document}
\begin{titlepage}
\samepage{
\setcounter{page}{1}
\rightline{LTH--656}
\rightline{\tt hep-th/0507229}
\rightline{July 2005}
\vfill
\begin{center}
 {\Large \bf  Higgs--Matter Splitting
\\
		in Quasi--Realistic Orbifold String GUTs\\}
\vfill
\vfill
 {\large Alon E. Faraggi\footnote{
        E-mail address: faraggi@amtp.liv.ac.uk}\\}
\vspace{.12in}
 {\it           Mathematical Sciences Department,
		University of Liverpool     
                Liverpool L69 7ZL\\
and\\
                Theoretical Physics Department,
		University of Oxford,
		Oxford OX1 3NP\\
and\\
                CERN, Theory Division, CH--1211 Geneva 23, Switzerland }
\end{center}
\vfill
\begin{abstract}
  {\rm
$E_6$ grand unification combines the Standard Model matter and Higgs
states in the single 27 representation. I discuss how the $E_6$ structure
underlies the quasi--realistic free fermion heterotic--string models.
$E_6\rightarrow SO(10)\times U(1)$ breaking is obtained by a GSO
phase in the $N=1$ partition function.
The equivalence of this symmetry breaking phase
with a particular choice of a boundary condition basis vector, which
is used in the quasi-realistic models, is demonstrated in several cases.
As a result matter states
in the spinorial 16 representation of $SO(10)$ arise from the twisted
sectors, whereas the Higgs states arise from the untwisted
sector. Possible additional phenomenological implications
of this $E_6$ symmetry breaking pattern are discussed.}
\end{abstract}
\smallskip}
\end{titlepage}

\setcounter{footnote}{0}

\def\l{\label}
\def\beq{\begin{equation}}
\def\eeq{\end{equation}}
\def\beqn{\begin{eqnarray}}
\def\eeqn{\end{eqnarray}}

\def\ie{{\it i.e.}}
\def\eg{{\it e.g.}}
\def\half{{\textstyle{1\over 2}}}
\def\third{{\textstyle {1\over3}}}
\def\quarter{{\textstyle {1\over4}}}
\def\m{{\tt -}}
\def\p{{\tt +}}

\def\slash#1{#1\hskip-6pt/\hskip6pt}
\def\slk{\slash{k}}
\def\GeV{\,{\rm GeV}}
\def\TeV{\,{\rm TeV}}
\def\y{\,{\rm y}}
\def\SM{Standard-Model }
\def\SUSY{supersymmetry }
\def\SSSM{supersymmetric standard model}
\def\vev#1{\left\langle #1\right\rangle}
\def\l{\langle}
\def\r{\rangle}

\def\Htw{{\tilde H}}
\def\chibar{{\overline{\chi}}}
\def\qbar{{\overline{q}}}
\def\ibar{{\overline{\imath}}}
\def\jbar{{\overline{\jmath}}}
\def\Hbar{{\overline{H}}}
\def\Qbar{{\overline{Q}}}
\def\abar{{\overline{a}}}
\def\alphabar{{\overline{\alpha}}}
\def\betabar{{\overline{\beta}}}
\def\tautwo{{ \tau_2 }}
\def\thetatwo{{ \vartheta_2 }}
\def\thetathree{{ \vartheta_3 }}
\def\thetafour{{ \vartheta_4 }}
\def\ttwo{{\vartheta_2}}
\def\tthree{{\vartheta_3}}
\def\tfour{{\vartheta_4}}
\def\ti{{\vartheta_i}}
\def\tj{{\vartheta_j}}
\def\tk{{\vartheta_k}}
\def\calF{{\cal F}}
\def\smallmatrix#1#2#3#4{{ {{#1}~{#2}\choose{#3}~{#4}} }}
\def\ab{{\alpha\beta}}
\def\Minv{{ (M^{-1}_\ab)_{ij} }}
\def\bone{{\bf 1}}
\def\ii{{(i)}}
\def\V{{\bf V}}
\def\b{{\bf b}}
\def\N{{\bf N}}
\def\t#1#2{{ \Theta\left\lbrack \matrix{ {#1}\cr {#2}\cr }\right\rbrack }}
\def\C#1#2{{ C\left\lbrack \matrix{ {#1}\cr {#2}\cr }\right\rbrack }}
\def\tp#1#2{{ \Theta'\left\lbrack \matrix{ {#1}\cr {#2}\cr }\right\rbrack }}
\def\tpp#1#2{{ \Theta''\left\lbrack \matrix{ {#1}\cr {#2}\cr }\right\rbrack }}
\def\l{\langle}
\def\r{\rangle}


\def\inbar{\,\vrule height1.5ex width.4pt depth0pt}

\def\IC{\relax\hbox{$\inbar\kern-.3em{\rm C}$}}
\def\IQ{\relax\hbox{$\inbar\kern-.3em{\rm Q}$}}
\def\IR{\relax{\rm I\kern-.18em R}}
 \font\cmss=cmss10 \font\cmsss=cmss10 at 7pt
\def\IZ{\relax\ifmmode\mathchoice
 {\hbox{\cmss Z\kern-.4em Z}}{\hbox{\cmss Z\kern-.4em Z}}
 {\lower.9pt\hbox{\cmsss Z\kern-.4em Z}}
 {\lower1.2pt\hbox{\cmsss Z\kern-.4em Z}}\else{\cmss Z\kern-.4em Z}\fi}

\def\AEF{A.E. Faraggi}
\def\NPB#1#2#3{{\it Nucl.\ Phys.}\/ {\bf B#1} (#2) #3}
\def\PLB#1#2#3{{\it Phys.\ Lett.}\/ {\bf B#1} (#2) #3}
\def\PRD#1#2#3{{\it Phys.\ Rev.}\/ {\bf D#1} (#2) #3}
\def\PRL#1#2#3{{\it Phys.\ Rev.\ Lett.}\/ {\bf #1} (#2) #3}
\def\PRP#1#2#3{{\it Phys.\ Rep.}\/ {\bf#1} (#2) #3}
\def\MODA#1#2#3{{\it Mod.\ Phys.\ Lett.}\/ {\bf A#1} (#2) #3}
\def\IJMP#1#2#3{{\it Int.\ J.\ Mod.\ Phys.}\/ {\bf A#1} (#2) #3}
\def\nuvc#1#2#3{{\it Nuovo Cimento}\/ {\bf #1A} (#2) #3}
\def\JHEP#1#2#3{{\it JHEP} {\textbf #1}, (#2) #3}
\def\etal{{\it et al\/}}

\hyphenation{su-per-sym-met-ric non-su-per-sym-met-ric}
\hyphenation{space-time-super-sym-met-ric}
\hyphenation{mod-u-lar mod-u-lar--in-var-i-ant}


\setcounter{footnote}{0}
\section{Introduction}
\bigskip
Grand Unification is well supported by the pattern of observed fermion and
gauge boson charges. Additionally, the observed logarithmic running of the
Standard Model parameters is compatible with the hypothesis of unification
in the gauge sector, and the heavy generation matter sector. Furthermore,
the longevity of the proton and the suppression of left--handed neutrino
masses also indicate a large unification scale of the order of $10^{16}GeV$.

Among the possible unification scenarios, $SU(5)$ is the most
economical. The observation of neutrino oscillations and
consequently of neutrino masses, necessitates adding $SU(5)$ singlets
and hence the need to go outside $SU(5)$. Matter unification in the
framework of $SO(10)$ is most compelling as it accommodates all the matter
states of a single generation in the 16 spinorial representation.
Then, a priori, one needs only two types of representations to accommodate the
Standard Model matter and Higgs spectrum,
the spinorial 16 and the vectorial 10 representations.
The framework of $E_6$ grand unification as even further appeal,
as, at the expense of adding an additional singlet, it embeds the
16 matter and 10 Higgs $SO(10)$ states into the 27 representation
of $E_6$ \cite{gutsreviews}.

As the observed symmetry at low energies consists solely of the
Standard Model symmetry, its embedding into a Grand Unification
group necessitates that we break the larger GUT symmetry. Grand
Unification introduces additional difficulties with proton decay
and neutrino masses. The Grand Unification gauge symmetry breaking
and the miscellanea issues typically require the introduction 
of large representations of the GUT gauge group, like the 126
of $SO(10)$ or the 351 of $E_6$, and devising complicated
symmetry breaking potentials to ensure proton longevity.

By producing a consistent framework for perturbative quantum gravity,
while simultaneously giving rise to gauge and matter structures,
string theory goes a step beyond conventional Grand Unified Theories
(GUTs). In the modern view of string theory, the different ten dimensional
string theories, as well as eleven dimensional supergravity, are effective
limits of a more fundamental theory, which at present is still unknown.
The heterotic limit \cite{hete},
in particular, gives rise to the Grand Unification
structures. Furthermore, the heterotic string is the only effective
limit that gives rise to spinorial representations in the perturbative 
spectrum, and hence is the only limit that can accommodate the $SO(10)$
and $E_6$ unification pictures \cite{canwit}. A class of string models that
accommodate the conventional GUT structures are the so--called
free fermionic models \cite{fsu5,fny,alr,eu,top,ks,lykken},
which are related to $Z_2\times Z_2$
orbifold compactification at special points in the moduli
space \cite{foc,fknr}.

String theory offers several additional advantages over conventional
GUTs. The replication of fermion families is associated with the
properties of the six dimensional compactified manifold. 
Depending on the properties of this internal manifold, string 
theory gives rise to novel gauge symmetry breaking mechanism,
which can be seen as breaking by GSO projections, or as 
breaking by Wilson line. Furthermore, string theory gives
rise to a doublet--triplet splitting mechanism, in which the color
triplets are projected out from the physical spectrum by GSO projections,
whereas the electroweak doublets remain. The GUT doublet--triplet splitting
problem then has a simple solution without the need to introduce
large representations. An explicit realization of the doublet--triplet
splitting in string GUT models was introduced in ref. \cite{ps}.

The doublet--triplet splitting is induced by the breaking
of the $SO(10)$ GUT to $SO(6)\times SO(4)$. The $SO(10)$
structure that underlies the three generation free fermionic models 
is well understood and has been amply exposed in the past.
However, the models in fact possess an underlying $E_6$ structure
that, for reasons explained here,
has been somewhat obscured in the past. It is the purpose
of this paper to remedy this situation and to expose the 
$E_6$ structure that underlies the realistic free fermionic
models. As discussed above, the characteristic feature of
$E_6$ is the unification of the matter and Higgs states
into the 27 representation of $E_6$. As is typical of string theory,
however, the $E_6$ symmetry is broken directly at the string level
by a GSO phase. As in the case of $SO(10)\rightarrow SO(6)\times SO(4)$,
the string induced breaking $E_6\rightarrow SO(10)\times U(1)$ has 
the additional consequence of projecting the twisted moduli \cite{moduli}, 
and may prove important for understanding the problem of supersymmetry
breaking.

\setcounter{footnote}{0}
\section{Realistic free fermionic models}

To elucidate the underlying $E_6$ structure of the realistic free 
fermionic models I discuss first the general structure of the
three generation models. 
In the free fermionic formulation \cite{fff} of the heterotic string
in four dimensions all the world--sheet
degrees of freedom  required to cancel
the conformal anomaly are represented in terms of free fermions
propagating on the string world--sheet.
In the light--cone gauge the world--sheet field content consists
of two transverse left-- and right--moving space--time coordinate bosons,
$X_{1,2}^\mu$ and ${\bar X}_{1,2}^\mu$,
and their left--moving fermionic superpartners $\psi^\mu_{1,2}$,
and additional 62 purely internal
Majorana--Weyl fermions, of which 18 are left--moving,
$\chi^{I}$, and 44 are right--moving, $\phi^a$.
In the supersymmetric sector the world--sheet supersymmetry is realized
non--linearly and the world--sheet supercurrent is given by
$T_F=\psi^\mu\partial X_\mu+i\chi^Iy^I\omega^I,~(I=1,\cdots,6).$
The $\{\chi^{I},y^I,\omega^I\}~(I=1,\cdots,6)$ are 18 real free
fermions transforming as the adjoint representation of $SU(2)^6$.
Under parallel transport around a noncontractible loop on the toroidal
world--sheet the fermionic fields pick up a phase
\begin{equation}
f~\rightarrow~-{\rm e}^{i\pi\alpha(f)}f~,~~\alpha(f)\in(-1,+1].
\label{fermionphase}
\end{equation}
A model in this construction \cite{fff}
is defined by a set of boundary conditions basis vectors
and by a choice of generalized GSO projection coefficients, which
satisfy the one--loop modular invariance constraints. The boundary
conditions basis vectors ${b}_k$ span a finite additive group
$
\Xi={\sum_k}n_i {b}_i
$
where $n_i=0,\cdots,{{N_{z_i}}-1}$.
The physical massless states in the Hilbert space of a given sector
$\alpha\in{\Xi}$ are then obtained by acting on the vacuum state of
that sector with the world-sheet bosonic and fermionic mode operators,
with frequencies $\nu_f$, $\nu_{f^*}$ and
by subsequently applying the generalized GSO projections,
\begin{equation}
\left\{e^{i\pi({b_i}F_\alpha)}-
{\delta_\alpha}c^*\left(\matrix{\alpha\cr
                 b_i\cr}\right)\right\}\vert{s}\rangle=0~,
\label{gsoprojections}
\end{equation}
where $F_\alpha(f)$ is a fermion number operator counting each mode of
$f$ once (and if $f$ is complex, $f^*$ minus once). For periodic
complex fermions [{\it i.e.} for $\alpha(f)=1)$]
the vacuum is a spinor in order to represent
the Clifford algebra of the corresponding zero modes.
For each periodic complex fermion $f$,
there are two degenerate vacua $\vert{+}\rangle$, $\vert{-}\rangle$,
annihilated by the zero modes $f_0$ and $f^*_0$ and with fermion
number $F(f)=0,-1$ respectively. In Eq. (\ref{gsoprojections}),
$\delta_\alpha=-1$ if $\psi^\mu$ is periodic in the sector $\alpha$,
and $\delta_\alpha=+1$ if $\psi^\mu$ is antiperiodic in the sector $\alpha$.

\subsection{An exemplary model}

The model in tables [\ref{nahe},\ref{m278}] provide an example of a three 
generation free fermionic model \cite{eu}. The model, the full massless
spectrum, and the trilevel superpotential are given in ref. \cite{eu}.
Various phenomenological aspects of this model were analyzed in the
literature \cite{eupheno}.

The boundary condition basis vectors which generate the realistic
free fermionic models are, in general, divided into two major subsets.
The first set consist of the NAHE set \cite{costas,nahe}, which is a set
of five boundary condition basis vectors denoted $\{{\bf 1},S,b_1,b_2,b_3\}$.
With `0' indicating Neveu-Schwarz (NS) boundary conditions
and `1' indicating Ramond boundary conditions, these vectors are as follows:

\beqn
 &&\begin{tabular}{c|c|ccc|c|ccc|c}
 ~ & $\psi^\mu$ & $\chi^{12}$ & $\chi^{34}$ & $\chi^{56}$ &
        $\bar{\psi}^{1,...,5} $ &
        $\bar{\eta}^1 $&
        $\bar{\eta}^2 $&
        $\bar{\eta}^3 $&
        $\bar{\phi}^{1,...,8} $ \\
\hline
\hline
      {\bf 1} &  1 & 1&1&1 & 1,...,1 & 1 & 1 & 1 & 1,...,1 \\
          $S$ &  1 & 1&1&1 & 0,...,0 & 0 & 0 & 0 & 0,...,0 \\
\hline
  ${b}_1$ &  1 & 1&0&0 & 1,...,1 & 1 & 0 & 0 & 0,...,0 \\
  ${b}_2$ &  1 & 0&1&0 & 1,...,1 & 0 & 1 & 0 & 0,...,0 \\
  ${b}_3$ &  1 & 0&0&1 & 1,...,1 & 0 & 0 & 1 & 0,...,0 \\
\end{tabular}
   \nonumber\\
   ~  &&  ~ \nonumber\\
   ~  &&  ~ \nonumber\\
     &&\begin{tabular}{c|cc|cc|cc}
 ~&      $y^{3,...,6}$  &
        $\bar{y}^{3,...,6}$  &
        $y^{1,2},\omega^{5,6}$  &
        $\bar{y}^{1,2},\bar{\omega}^{5,6}$  &
        $\omega^{1,...,4}$  &
        $\bar{\omega}^{1,...,4}$   \\
\hline
\hline
    {\bf 1} & 1,...,1 & 1,...,1 & 1,...,1 & 1,...,1 & 1,...,1 & 1,...,1 \\
    $S$     & 0,...,0 & 0,...,0 & 0,...,0 & 0,...,0 & 0,...,0 & 0,...,0 \\
\hline
${b}_1$ & 1,...,1 & 1,...,1 & 0,...,0 & 0,...,0 & 0,...,0 & 0,...,0 \\
${b}_2$ & 0,...,0 & 0,...,0 & 1,...,1 & 1,...,1 & 0,...,0 & 0,...,0 \\
${b}_3$ & 0,...,0 & 0,...,0 & 0,...,0 & 0,...,0 & 1,...,1 & 1,...,1 \\
\end{tabular}
\label{nahe}
\eeqn
with the following
choice of phases which define how the generalized GSO projections are to
be performed in each sector of the theory:
\beq
      c\left( \matrix{b_i\cr b_j\cr}\right)~=~
      c\left( \matrix{b_i\cr S\cr}\right) ~=~
     -c\left( \matrix{\bone \cr \bone \cr}\right) ~= ~ -1~.
\label{nahephases}
\eeq
The NAHE set [\ref{nahe}] is common subset to all the models discussed here,
and therefore will be dropped in the following. 
The gauge group at the level of the NAHE set is 
$$SO(10)\times SO(6)^3\times E_8~. $$
The $SO(10)$ group gives rise to the universal part of the observable
gauge group. The $SO(6)$ groups are flavor dependent symmetries, while
the $E_8$ group is hidden, as the Standard Model states are neutral
under this group. The NAHE--set basis vectors $b_1$, $b_2$ and $b_3$
correspond to the three twisted sectors of the $Z_2\times Z_2$ orbifold.
At the level of the NAHE--set the free fermionic models contain 48 chiral
generations and correspond to a so--called ``orbifold string GUT''.

To reduce the number of generations and break the
GUT symmetry one introduces three additional
basis vectors, typically denoted as $\alpha$, $\beta$ and $\gamma$.
The additional basis vectors that generate the string model of ref.
\cite{eu} are displayed in table [\ref{m278}].
\beqn
 &\begin{tabular}{c|c|ccc|c|ccc|c}
 ~ & $\psi^\mu$ & $\chi^{12}$ & $\chi^{34}$ & $\chi^{56}$ &
        $\bar{\psi}^{1,...,5} $ &
        $\bar{\eta}^1 $&
        $\bar{\eta}^2 $&
        $\bar{\eta}^3 $&
        $\bar{\phi}^{1,...,8} $\\
\hline
\hline
  ${\alpha}$  &  0 & 0&0&0 & 1~1~1~0~0 & 0 & 0 & 0 & 1~1~1~1~0~0~0~0 \\
  ${\beta}$   &  0 & 0&0&0 & 1~1~1~0~0 & 0 & 0 & 0 & 1~1~1~1~0~0~0~0 \\
  ${\gamma}$  &  0 & 0&0&0 &
        $\frac{1}{2}$~$\frac{1}{2}$~$\frac{1}{2}$~$\frac{1}{2}$~$\frac{1}{2}$
          & $\frac{1}{2}$ & $\frac{1}{2}$ & $\frac{1}{2}$ &
                $\frac{1}{2}$~0~1~1~$\frac{1}{2}$~$\frac{1}{2}$~$\frac{1}{2}$~0
\\
\end{tabular}
   \nonumber\\
   ~  &  ~ \nonumber\\
   ~  &  ~ \nonumber\\
     &\begin{tabular}{c|c|c|c}
 ~&   $y^3{y}^6$
      $y^4{\bar y}^4$
      $y^5{\bar y}^5$
      ${\bar y}^3{\bar y}^6$
  &   $y^1{\omega}^5$
      $y^2{\bar y}^2$
      $\omega^6{\bar\omega}^6$
      ${\bar y}^1{\bar\omega}^5$
  &   $\omega^2{\omega}^4$
      $\omega^1{\bar\omega}^1$
      $\omega^3{\bar\omega}^3$
      ${\bar\omega}^2{\bar\omega}^4$ \\
\hline
\hline
$\alpha$ & 1 ~~~ 0 ~~~ 0 ~~~ 0  & 0 ~~~ 0 ~~~ 1 ~~~ 1  & 0 ~~~ 0 ~~~ 1 ~~~ 1
\\
$\beta$  & 0 ~~~ 0 ~~~ 1 ~~~ 1  & 1 ~~~ 0 ~~~ 0 ~~~ 0  & 0 ~~~ 1 ~~~ 0 ~~~ 1
\\
$\gamma$ & 0 ~~~ 1 ~~~ 0 ~~~ 1  & 0 ~~~ 1 ~~~ 0 ~~~ 1  & 1 ~~~ 0 ~~~ 0 ~~~ 0
\\
\end{tabular}
\label{m278}
\eeqn
\beq
      c\left( \matrix{b_i\cr \alpha,\beta, \gamma\cr}\right)=
     -c\left( \matrix{\alpha,\beta\cr \bone\cr}\right) =
      c\left( \matrix{\alpha \cr \beta \cr}\right)  =
      c\left( \matrix{\gamma \cr \alpha \cr}\right) =
     -c\left( \matrix{\gamma \cr \beta \cr}\right) 
      =  -1~,~(i=1,2,3),
\label{m278phases}
\eeq
with the others specified by modular invariance and space--time
supersymmetry. The boundary condition basis vector in [\ref{m278}]
break the gauge group to:
$$SU(3)\times SU(2)\times U(1)^2~~\times ~~U(1)^6 ~~\times SU(5)\times 
SU(3)\times U(1)^2,$$
where the first two $U(1)$s arise from the $SO(10)$ group, the next six
$U(1)$s are obtained from $SO(6)^3$, and the remaining two $U(1)$
arise from the hidden $E_8$ gauge group. Additionally, the 
basis vectors $\alpha$, $\beta$, $\gamma$ reduce the number
of generations to three. One from each of the twisted sectors
$b_1$, $b_2$ and $b_3$. Electroweak Higgs doublets are obtained
from the untwisted sector, and the sector $b_1+b_2+\alpha+\beta$. 
The full spectrum of this model, and detailed phenomenological
studies are given in the literature \cite{eu,eupheno}.

{}From the above we see that the model exhibits an underlying $SO(10)$
symmetry, but there is no trace of an $E_6$ group. It is the purpose
of this paper to elucidate the stringy $E_6 \rightarrow SO(10)\times U(1)$
breaking, and to show that, just as in the case of the stringy
$SO(10) \rightarrow SO(6)\times SO(4)$ the stringy breaking of $E_6$
has additional phenomenological consequences.

\setcounter{footnote}{0}
\section{$E_6$ origins}
To expose the underlying $E_6$ structure of the free fermionic models, 
we have to look at subsets of the basis vectors, or of the partition,
function that preserve the $E_6$ symmetry. A good starting 
point is the subset of basis vectors 
\beq
\{\bone, S, 2\gamma, \xi_2=\bone+b_1+b_2+b_3\}~.
\label{2gammaneq4}
\eeq
This subset generates an $N=4$ SUSY vacuum with $SO(12)\times SO(16)\times
SO(16)$ gauge group. The NS sector gives rise to the space--time vector
bosons that generate $SO(12)\times SO(16)\times SO(8)\times SO(8)$, and 
the sector $\xi_2$ complements the hidden gauge group to $SO(16)$ \cite{foc}.
Adding the basis vectors $b_1$ and $b_2$ then breaks $N=4$ to $N=1$
supersymmetry. It breaks the gauge group to
$SO(4)^3\times SO(10)\times U(1)^3\times SO(16)$, and introduces 24
observable matter multiplets in the spinorial 16 representation of the
observable $SO(10)$, from the sectors $b_1$, $b_2$ and $b_3=\bone+b_1+b_2+
\xi_2$, and 16 hidden matter multiplets in the vectorial
16 representation of the hidden $SO(16)$ gauge group from
the sectors $b_j+(2\gamma\oplus\xi_2)$ $j=1,2,3$. The symbol $\oplus$
is used here to indicate that there are two sectors that produce the
hidden matter representations. One being $b_j+2\gamma$ and the second
$b_j+2\gamma+\xi_2$. This notation will be used in the following.

Note that we could have projected the enhancing vector bosons from the sector
$\xi_2$  by the choice of GSO phase $c{\xi_2\choose S}=-1$, where $S$ is the
SUSY generator. The price is that the SUSY generators are projected and
the vacuum is tachyon free and nonsupersymmetric. The reason that there are
no tachyon is because the only tachyons in the model arise from the NS
sector, and the projections of those only depend on $\delta_S$, and not
on the phase $c{\xi_2\choose S}$. This is reminiscent of the ten dimensional
heterotic $SO(16)\times SO(16)$ model, in which modular invariance forces
that the GSO phase that breaks $E_8\times E_8\rightarrow SO(16)\times SO(16)$,
also projects out the space--time supersymmetry. 

So far there is no reminiscence of $E_6$. An alternative way to produce
the model of (\ref{2gammaneq4}) is by starting with the set of basis
vectors 
\beq
\{\bone,S,\xi_1,\xi_2=\bone+b_1+b_2+b_3\}
\label{xi1xi2}
\eeq
with
\beq
\xi_1=(0,\cdots,0\vert{\underbrace{1,\cdots,1}_{{\bar\psi^{1,\cdots,5}},
{\bar\eta^{1,2,3}}}},0,\cdots,0)~.
\label{vectorxi1}
\eeq
for a suitable choice of GSO phases,
this set generates an $N=4$ vacuum. The four dimensional
gauge group in this model depends on the discrete choice of
the GSO phase $$c{\xi_1\choose \xi_2}=\pm1~.$$
Since the overlap of periodic fermions between $\xi_1$ and $\xi_2$ is empty,
we note from (\ref{gsoprojections}) that the choice 
$c{\xi_1\choose \xi_2}=-1$ projects all the states from
the sectors $\xi_1$ and $\xi_2$, whereas the choice 
$c{\xi_1\choose \xi_2}=+1$ retains them in the spectrum.
Thus, the choice 
\beq
c{\xi_1\choose \xi_2}=+1
\label{cxi1xi2plus}
\eeq
produces a
model with $SO(12)\times E_8\times E_8$ gauge group, 
whereas the choice 
\beq
c{\xi_1\choose \xi_2}=-1
\label{cxi1xi2minus}
\eeq
produces a model with $SO(12)\times SO(16)\times SO(16)$
gauge group, and reproduces the spectrum of (\ref{2gammaneq4}).
Thus, we note that there are two distinct ways to
generate the same model. One is by the mapping $\xi_1\rightarrow 2\gamma$,
and the alternative method by the choice of the discrete phase
$c{\xi_1\choose \xi_2}$.

Adding the basis vectors $\{b_1,b_2\}$ to the set (\ref{xi1xi2})
corresponds to the $Z_2\times Z_2$ orbifold projection.  
This breaks $N=4$ to $N=1$ supersymmetry. Setting
$c{\xi_1\choose\xi_2}=+1$ generates the $SO(4)^3\times 
E_6\times U(1)^2\times E_8$. The sectors $b_j$ consist of 
12 Ramond fermions and produce states in
the spinorial 16 representation of $SO(10)$, whereas the 
sectors $b_j+\xi_1$ produce a matching number of states in the
vectorial 10 representation of $SO(10)$. In addition the sectors
$b_j+\xi_1$ produce a matching number of $SO(10)$ singlets
which are charged under the $U(1)$ in the decomposition
$E_6\rightarrow SO(10)\times U(1)$ and a matching number
of $E_6$ singlets. The untwisted NS sector produces
6 vectorial 10 multiplets, a matching number of 
$SO(10)$ singlets, and a matching number of $E_6$
singlets. The sector $\xi_1$ produces 3 $16$ multiplets 
and 3 $\overline{16}$ multiplets. Thus, in this case
we get a model with 24 multiplets in the $27$ representation
of $E_6$ from the twisted sectors and 3 pairs in the 
$27+\overline{27}$ from the untwisted sector.

Setting $c{\xi_1\choose\xi_2}=-1$ projects the vector bosons
from the sectors $\xi_1$ and $\xi_2$. Vector bosons therefore
are obtained solely from the untwisted sector, which produces
the $SO(16)\times SO(16)$ gauge group. 
Adding the $Z_2\times Z_2$ twists breaks the gauge group to
$SO(4)^3\times SO(10)\times U(1)^3\times SO(16)$.
The twisted sectors $b_j$ still produce the 24 multiplets in the
spinorial 16 representation of $SO(10)$, but now the sectors
$b_j+\xi_1$ produce states in the vectorial 16 representation 
of the hidden $SO(16)$ gauge group. The same spectrum
is reproduced by replacing the basis vector $\xi_1$ with the 
basis vector $2\gamma$. In this case the overlap between
$\xi_2$ and $2\gamma$ is not empty. Therefore, the projection
${\xi_2\choose{2\gamma}}$ cannot project all the states from $\xi_2$,
but merely halves the spectrum from this sector, whereas the sector 
$2\gamma$ does not produce massless states. The basis vectors
$b_1$ and $b_2$ reduce the number of supersymmetries and 
break the gauge symmetry as before. The matter multiplets from the untwisted
sector and the twisted sectors $b_j$ remain as before, and the sectors
$b_j+(2\gamma\oplus\xi_2)$
now produce the 24 vectorial 16 representations
of the hidden $SO(16)$ gauge group.

It is therefore noted that the map 
\beq
\xi_1\rightarrow 2\gamma~,
\label{xi1to2gamma}
\eeq
is in fact equivalent to the discrete choice of GSO phase
\beq
c{\xi_1\choose \xi_2}=+1\rightarrow c{\xi_1\choose \xi_2}=-1,
\label{cxi1xi2tominus}
\eeq
and that the later corresponds to the 
gauge symmetry breaking pattern $E_6\rightarrow SO(10)\times U(1)$.
It is noted that the map (\ref{xi1to2gamma}) also requires the
phase map
\beq
c{\xi_1\choose \xi_1}\rightarrow -c{{2\gamma}\choose{2\gamma}}~.
\label{xi1xi12g2g}
\eeq
So far I discussed the models only at the $N=4$ level and at the 
$N=1$ $Z_2\times Z_2$ orbifold level. In the following I turn to examine
how this structure is manifested in the case of quasi--realistic three
generation models. In this regard it should be noted that the original
construction of three generation free fermion models, that utilize
the NAHE--subset of basis vectors, obscures the underlying $E_6$
structure of these models. The reason is that these models utilize
the vector $\gamma$ to break the observable 
$SO(2n)\rightarrow SU(n)\times U(1)$. The vector $2\gamma$, which
separates the gauge degrees of freedom from the geometrical degrees
of freedom, therefore arises only as a multiple of the vector $\gamma$.
The vector $2\gamma$ also fixes the charges of the chiral generations
under $U(1)$'s in the $E_8$ Cartan subalgebra, which are external to $E_6$, 
and hence reduces the NAHE--base generations by 1/2. Thus, the NAHE--base,
supplemented with the $2\gamma$, or $\xi_1$, contains 24 chiral generations, 
as opposed to the NAHE--set by itself, which contains 48 chiral generations. 
The remaining reduction to three generations is obtained by the action of
the basis vectors $\{\alpha,\beta, \gamma\}$ on the internal free fermions
$\{y,\omega|{\bar y},{\bar\omega}\}^{1,\cdots,6}$, each inducing a 
$Z_2$ projection on each of the twisted sectors $b_j$ $(j=1,2,3,)$.
Hence, reducing the number of generations in each from eight to one.
Models that do not contain the vector $\gamma$, like $SO(6)\times SO(4)$
models, must explicitly include the vector $2\gamma$, or $\xi_1$,
in the basis, to reduce the number of generations to three.

The model of table [\ref{toyso64model}] is constructed
to study the map (\ref{xi1to2gamma}) in a quasi--realistic model. 
It should be emphasized that the aim is not to construct a 
realistic model, but merely to study the map in a model
that shares some of the structure of the three generation free
fermionic models. In particular the assignment of boundary
conditions with respect to the internal world--sheet fermions
$\{y,\omega\vert{\bar y},{\bar\omega}\}$ is reminiscent of
this assignment in the three generation free fermionic models.
The model in table [\ref{toyso64model}] is generated
by the subset of basis vectors $\{\bone, S, \xi_1,\xi_2,b_1,b_2\}$,
and the additional basis vectors $\{b_4,b_5,\alpha\}$ in table
[\ref{toyso64model}].

\beqn
 &\begin{tabular}{c|c|ccc|c|ccc|c}
 ~ & $\psi^\mu$ & $\chi^{12}$ & $\chi^{34}$ & $\chi^{56}$ &
        $\bar{\psi}^{1,...,5} $ &
        $\bar{\eta}^1 $&
        $\bar{\eta}^2 $&
        $\bar{\eta}^3 $&
        $\bar{\phi}^{1,...,8} $ \\
\hline
\hline
  ${\alpha}$  &  1 & 1&0&0 & 1~1~1~1~1 & 1 & 0 & 0 & 0~0~0~0~0~0~0~0 \\
  ${\beta}$   &  1 & 0&1&0 & 1~1~1~1~1 & 0 & 1 & 0 & 0~0~0~0~0~0~0~0 \\
  ${\gamma}$  &  0 & 0&0&0 & 1~1~1~0~0 & 1 & 1 & 1 & 0~0~1~1~0~0~0~0 \\
\end{tabular}
   \nonumber\\
   ~  &  ~ \nonumber\\
   ~  &  ~ \nonumber\\
     &\begin{tabular}{c|c|c|c}
 ~&   $y^3{\bar y}^3$
      $y^4{\bar y}^4$
      $y^5{\bar y}^5$
      $y^6{\bar y}^6$
  &   $y^1{\bar y}^1$
      $y^2{\bar y}^2$
      $\omega^5{\bar\omega}^5$
      $\omega^6{\bar\omega}^6$
  &   $\omega^1{\bar\omega}^1$
      $\omega^2{\bar\omega}^2$
      $\omega^3{\bar\omega}^3$
      $\omega^4{\bar\omega}^4$ \\
\hline
\hline
$\alpha$ & 1 ~~~ 0 ~~~ 0 ~~~ 1  & 0 ~~~ 0 ~~~ 1 ~~~ 0  & 0 ~~~ 0 ~~~ 0 ~~~ 1 \\
$\beta$  & 0 ~~~ 0 ~~~ 0 ~~~ 1  & 0 ~~~ 1 ~~~ 1 ~~~ 0  & 1 ~~~ 0 ~~~ 0 ~~~ 0 \\
$\gamma$ & 0 ~~~ 1 ~~~ 0 ~~~ 1  & 0 ~~~ 1 ~~~ 0 ~~~ 1  & 0 ~~~ 1 ~~~ 0 ~~~ 1 \\
\end{tabular}
\label{toyso64model}
\eeqn
With the choice of generalized GSO coefficients:
\beqn
c\left(\matrix{S\cr
                                    a_j\cr}\right)=\delta_{a_j}~~~,~~~
&& c\left(\matrix{b_{1,2,4,5}\cr
                  b_{1,2,4,5},\xi_1,\xi_2,\alpha\cr}\right)=
c\left(\matrix{{\bone}\cr
                          \xi_1,\xi_2\cr}\right)=~-1\nonumber\\
&& c\left(\matrix{\bone\cr
                                   \alpha\cr}\right)=
c\left(\matrix{\xi_1\cr
                                    {\xi_2},\alpha\cr}\right)=~1~,\nonumber
\eeqn
with the others specified by modular
invariance and space--time supersymmetry.
The gauge group of the model arises as follows.
The NS sector produces the generators of the
observable and hidden gauge groups
\beq
(SO(6)\times SU(2)_L\times SU(2)_R
\times U(1)_{1,2,3})_{\rm O}~
\times~(SO(12)\times SU(2)_{H_1}\times SU(2)_{H_2})_{\rm H},
\label{nsvectorbosons}
\eeq 
and the sectors $\xi_1$, and $\xi_2$, enhance the observable, and hidden,
gauge groups of the model, respectively to:
\beqn
{\rm observable}~: && SU(6)\times SU(2)_L\times U(1)^2~,\\
    {\rm hidden}~: && E_7\times SU(2)~,
\eeqn
where $SU(2)_R$ and the $U(1)$ combination,
\beq
U(1)_6^\prime~=~ U(1)_1+U(1)_2-U(1)_3~,
\label{u1p6}
\eeq
are embedded in $SU(6)$, and the two orthogonal $U(1)$ combinations
are given by
\beqn
& U(1)_1^\prime~=~ U(1)_1-U(1)_2~,\\
& U(1)_2^\prime~=~ U(1)_1+U(1)_2+2U(1)_3~.
\label{u1p2u1p3}
\eeqn
Similarly, the hidden $SO(12)\times SU(2)_{H_1}$ are enhanced
by the states from the sector $\xi_2$ to produce the $E_7$
gauge group.

This model is not a realistic model. It preserves some of the 
structure of the quasi--realistic string models in the sense that 
it produces three chiral generations from the sectors, $b_1$, $b_2$
and $b_3$. But the full spectrum is not realistic as it contains
additional chiral matter, and the untwisted electroweak Higgs are
projected out. The purpose here is to
study how the maps (\ref{xi1to2gamma}) and (\ref{cxi1xi2tominus}) are
related in a model that preserves some of this realistic structure.
In the model of (\ref{toyso64model}), with the choice of phases above,
the sectors $b_j\oplus b_j+ \xi_1$ produce three chiral generations
in the $(15,1)+(6,2)$ of $SU(6)\times SU(2)_L$. The sectors 
$b_{4,5}\oplus b_{4,5}+\xi_1$ and $b_4+b_5\oplus b_4+b_5+\xi_1$
produce states in the $(15,1)+({\bar 6},2)$ and the sectors
containing $\alpha$, which breaks the NS $SO(10)$ gauge subgroup,
produce states that transform as $(6,1)+(1,2)$ under $SU(6)\times SU(2)_L$
and transform as doublets under the hidden $SU(2)$ gauge group. 
These sectors are: 
$b_2+b_4+\alpha$,
$b_1+b_4+b_5+\alpha$,
$b_1+b_2+b_4+\alpha$,
$b_2+b_3+b_5+\alpha$,
$b_3+b_5+\alpha$, where I heuristically defined 
the combination $b_3=\bone+b_1+b_2+\xi_2$. The full massless spectrum of this
model is given in appendix \ref{modelxi1so10spec}. Table [\ref{xi1xi2plus}]
contains the states in this model that originate from sectors that
preserve the $SO(10)$ symmetry of the NS sector, whereas
table [\ref{xi1xi2pluse}] contains the $SO(10)$ breaking spectrum. 

We can now project the vector bosons from the sectors $\xi_1$ and $\xi_2$
by the map (\ref{cxi1xi2tominus}) which fixes the phase (\ref{cxi1xi2minus}).
The full massless spectrum of this model is given in appendix
\ref{modelxi2so10spec}, where tables [\ref{xi1xi2minus}],
and [\ref{xi1xi2minuse}],
contain the $SO(10)$ preserving, and $SO(10)$ breaking, spectrum, 
respectively. The GSO projections now project the states from the
sectors $\xi_1$ and $\xi_2$. The gauge group in this case 
arises solely from the NS sector and the four dimensional gauge 
group is that of eq. (\ref{nsvectorbosons}). In this case the
sectors $b_j\oplus\xi_1$ split. The sectors $b_j$ produce spinorial matter
states of the observable gauge group, whereas the sectors $b_j+\xi_1$
produce vectorial matter states of the hidden gauge group.
Thus, the would be twisted Higgs states are projected from 
the physical spectrum by this splitting. Similarly, it is noted 
from table [\ref{xi1xi2minus}] that the spinorial matter states
from the sector $S+b_4+b_5$ are projected out from the physical
spectrum, and this sector produces vectorial matter states
in the observable sector. Therefore, the original $E_6$ embedding
of the spinorial (or matter) and vectorial (or Higgs), representations,
which is ``remembered'' in the $SU(6)$, $(15,1)$ and $(6,2)$, 
representations, is broken by the choice of GSO projection phase,
eq. (\ref{cxi1xi2minus}). Similarly, the $SO(10)$ breaking spectrum
in this model, shown in table [\ref{xi1xi2minuse}], is split between
the sectors that contain, and do not contain, $\xi_1$, which in 
the previous model were combined. 

We can now perform the map (\ref{xi1to2gamma}). Since the overlap 
between $\xi_2$ and $2\gamma$ is now not empty, we can choose
the phase $c{\xi_1\choose{2\gamma}}=-1$. Choosing the opposite
phase amounts to redefinition of the charges, and has no physical
effect. In the hidden sector of this model the NS sector generate
the gauge group
\beq
(SO(8)\times SU(2)_{H_3}\times SU(2)_{H_4}\times 
SU(2)_{H_1}\times SU(2)_{H_2})_{\rm H},
\label{nshiddenvectors}
\eeq
and the sector $\xi_2$ enhances the hidden sector gauge symmetry to
$SO(12)\times SU(2)_{H_3}\times SU(2)_{H_4}$, which is identical
to that of model \ref{modelxi2so10spec}. Thus, in this 
model the vectorial hidden sector matter states arise from the sectors
$b_j+2\gamma$ and $b_j+2\gamma+\xi_2$. 
The symbol $\oplus$ is used to indicate this combination
in table [\ref{xi2gammae}].
Inspecting the spectrum in appendices
\ref{modelxi2so10spec} and \ref{model2gso10spec}, we note that
the spectrum is indeed identical with the map
\beq
\xi_1\leftrightarrow (2\gamma\oplus\xi_2)~.
\label{xi1to2gammapxi2map}
\eeq
Note that in table [\ref{xi2gammae}] the states arise from separate
sectors and do not combine, as is the case in table [\ref{xi1xi2minuse}]
of model \ref{modelxi2so10spec}.

A similar map operates in models which utilize the vector
$\gamma$ and hence break the $SO(10)$ symmetry to $SU(5)\times U(1)$
or to $SU(3)\times SU(2)\times U(1)^2$. Supplementing the NAHE--set
basis vectors with the set of basis vectors in table
[\ref{fsu51}]
\beqn
 &\begin{tabular}{c|c|ccc|c|ccc|c}
 ~ & $\psi^\mu$ & $\chi^{12}$ & $\chi^{34}$ & $\chi^{56}$ &
        $\bar{\psi}^{1,...,5} $ &
        $\bar{\eta}^1 $&
        $\bar{\eta}^2 $&
        $\bar{\eta}^3 $&
        $\bar{\phi}^{1,...,8} $ \\
\hline
\hline
  ${\alpha}$  &  0 & 0&0&0 & 1~1~1~1~1 & 1 & 0 & 0 & 0~0~0~0~0~0~0~0 \\
  ${\beta}$   &  0 & 0&0&0 & 1~1~1~1~1 & 0 & 1 & 0 & 0~0~0~0~0~0~0~0 \\
  ${\gamma^\prime}$
              &  0 & 0&0&0 &
		${1\over2}$~${1\over2}$~${1\over2}$~${1\over2}$~${1\over2}$
	      & ${1\over2}$ & ${1\over2}$ & ${1\over2}$ &
                0~0~0~0~0~1~1~1 \\
\end{tabular}
   \nonumber\\
   ~  &  ~ \nonumber\\
   ~  &  ~ \nonumber\\
     &\begin{tabular}{c|c|c|c}
 ~&   $y^3{y}^6$
      $y^4{\bar y}^4$
      $y^5{\bar y}^5$
      ${\bar y}^3{\bar y}^6$
  &   $y^1{\omega}^5$
      $y^2{\bar y}^2$
      $\omega^6{\bar\omega}^6$
      ${\bar y}^1{\bar\omega}^5$
  &   $\omega^2{\omega}^4$
      $\omega^1{\bar\omega}^1$
      $\omega^3{\bar\omega}^3$
      ${\bar\omega}^2{\bar\omega}^4$ \\
\hline
\hline
$b_4$    & 1 ~~~ 0 ~~~ 0 ~~~ 1  & 0 ~~~ 0 ~~~ 1 ~~~ 1  & 0 ~~~ 0 ~~~ 1 ~~~ 1 \\
$b_5$    & 0 ~~~ 0 ~~~ 1 ~~~ 1  & 1 ~~~ 0 ~~~ 0 ~~~ 1  & 0 ~~~ 1 ~~~ 0 ~~~ 1 \\
$\gamma^\prime$ 
         & 0 ~~~ 1 ~~~ 0 ~~~ 0  & 0 ~~~ 1 ~~~ 0 ~~~ 0  & 1 ~~~ 0 ~~~ 0 ~~~ 0 \\
\end{tabular}
\label{fsu51}
\eeqn
and the choice of the additional generalized GSO coefficients:
\beqn
&& c\left(\matrix{\bone\cr
                           b_4,b_5,\gamma^\prime\cr}\right)=
c\left(\matrix{b_4,b_5,\gamma^\prime\cr
                           b_1,b_2,b_3\cr}\right)=
c\left(\matrix{\gamma^\prime\cr
                                    b_4,b_5\cr}\right)=
-c\left(\matrix{b_4\cr
                                    b_5\cr}\right)=+1\nonumber\\
&&
 c\left(\matrix{S\cr
                                   a_j\cr}\right)=\delta_{a_j}\nonumber
\eeqn

The gauge group of this model is:
\beq
(SU(5)\times U(1)\times U(1)_{1,2,3}\times U(1)_{4,5,6})_{\rm O}\times 
(SO(10)\times SO(6))_{\rm H}~,
\label{ggfsu5model}
\eeq
where $U(1)_{1,2,3}$ are embedded in the observable $E_8$,
whereas $U(1)_{4,5,6}$ are from the complexified world-sheet
real fermions,
$({\bar y}^3+i{\bar y}^6)$,
$({\bar y}^1+i{\bar\omega}^5)$,
$({\bar\omega}^2+i{\bar\omega}^4)$.
Space--time vector bosons in this model arise solely from
the Neveu--Schwarz sector. The model contains several additional 
combinations of basis vectors that may a priori give rise to
vector bosons. These include the sectors: $2\gamma^\prime$,
$\xi_2=\bone+b_1+b_2+b_3$ and the sector
$S+b_1+b_2+b_3+b_4+b_5\pm\gamma^\prime$.
All the states from these sectors are projected out and hence
there is no enhancement of the NS gauge group in this model.
In the case of the sector $2\gamma^\prime$ the projection depends
on $\gamma\cdot b_j$. If $\gamma^\prime\cdot b_j={\rm even}$ \&
 $\gamma^\prime\cdot b_i={\rm odd}$, with $i\ne j$, vector bosons
from $2\gamma^\prime$ are projected out. This may occur because
$\gamma^\prime$ must contain periodic internal fermions from
the set $\{y,\omega\vert{\bar y},{\bar\omega}\}$. The reason
is that $\gamma^\prime$ breaks the $SO(10)$ symmetry and simultaneously
halves the number of generations by fixing the $U(1)_{1,2,3}$ charges. 
In models with only periodic boundary conditions the later function is
performed by the vector $2\gamma$, which does not break the
gauge group. Thus, in NAHE--based models with 1/2 boundary conditions,
we must assign in $\gamma$ periodic boundary conditions to internal fermions
to insure that full $SO(10)$ spinorial 16 representations remain in the
physical spectrum. This means that we have the freedom to choose
appropriate boundary conditions that project the vector bosons from
$2\gamma^\prime$. Additionally, with the choice of phases above
the vector bosons from the sector $S+b_1+b_2+b_3+b_4+b_5\pm\gamma^\prime$
are projected out. 

The model of [\ref{fsu51}] then contains three generations of $SO(10)$
chiral 16 representations, decomposed under $SU(5)\times U(1)$ from the 
sectors $b_{1,2,3}$; three generations of the hidden $SO(16)$ vectorial
16 representation, decomposed under the hidden $SO(10)\times SO(6)$
gauge group, The sectors $b_2+b_5$, $b_1+b_4$ and $S+b_1+b_2+b_4+b_5$
produce states that are $E_8\times E_8$ singlets, and are charged with
respect to the complexified internal fermions,
$\{ {\bar y}^3{\bar y}^6;
{\bar y}^1{\bar\omega}^5; 
{\bar\omega}^2{\bar\omega}^4\}$. The sectors 
$b_3\pm\gamma^\prime$;
$S+b_2+b_3+b_5\pm\gamma^\prime$;
$S+b_2+b_3+b_4+b_5\pm\gamma^\prime$;
$S+b_1+b_3+b_4\pm\gamma^\prime$;
$S+b_1+b_3+b_4+b_5\pm\gamma^\prime$;
$S+b_1+b_2+b_4+b_5\pm\gamma^\prime$,
produce fractionally charged matter
states that transform as $4+{\bar 4}$ of the hidden $SO(6)$ gauge
group. Note that in this model the entire set of untwisted geometrical 
moduli are projected out due to the specific pairing of the left--moving
real fermions into complex pairs \cite{moduli}. Additionally, the twisted
moduli from the sectors $b_{1,2,3}$ are projected out as well \cite{moduli}.
The NS sector in this model produces, in addition to the gravity and
gauge multiplets, scalar states that are charged with respect to
$U(1)_{4,5,6}$. 
I note that this is not a realistic model as it does not contain the
Higgs representations that are needed to break the GUT and electroweak
symmetries. 

I now turn to show how the map (\ref{xi1to2gamma}) is implemented
in this model. The map is induced by the substitution
$\gamma^\prime \rightarrow \gamma$, with $\gamma$ given
in table [\ref{newgamma}]
\beqn
 &\begin{tabular}{c|c|ccc|c|ccc|c}
 ~ & $\psi^\mu$ & $\chi^{12}$ & $\chi^{34}$ & $\chi^{56}$ &
        $\bar{\psi}^{1,...,5} $ &
        $\bar{\eta}^1 $&
        $\bar{\eta}^2 $&
        $\bar{\eta}^3 $&
        $\bar{\phi}^{1,...,8} $ \\
\hline
\hline
  ${\gamma}$  &  0 & 0&0&0 &
		${1\over2}$~${1\over2}$~${1\over2}$~${1\over2}$~${1\over2}$
	      & ${1\over2}$ & ${1\over2}$ & ${1\over2}$ &
                ${1\over2}$~${1\over2}$~${1\over2}$~${1\over2}$~0~0~1~1 \\
\end{tabular}
   \nonumber\\
   ~  &  ~ \nonumber\\
   ~  &  ~ \nonumber\\
     &\begin{tabular}{c|c|c|c}
 ~&   $y^3{y}^6$
      $y^4{\bar y}^4$
      $y^5{\bar y}^5$
      ${\bar y}^3{\bar y}^6$
  &   $y^1{\omega}^5$
      $y^2{\bar y}^2$
      $\omega^6{\bar\omega}^6$
      ${\bar y}^1{\bar\omega}^5$
  &   $\omega^2{\omega}^4$
      $\omega^1{\bar\omega}^1$
      $\omega^3{\bar\omega}^3$
      ${\bar\omega}^2{\bar\omega}^4$ \\
\hline
\hline
$\gamma$ & 0 ~~~ 1 ~~~ 0 ~~~ 0  & 0 ~~~ 1 ~~~ 0 ~~~ 0  & 1 ~~~ 0 ~~~ 0 ~~~ 0 \\
\end{tabular}
\label{newgamma}
\eeqn
Additionally, modular invariance requires the phase modification
\beq
c{\gamma\choose\gamma}=-c{\gamma^\prime\choose\gamma^\prime}~.
\label{gammapgammaphasemodification}
\eeq
All other GSO phases are identical in the two models. The gauge
group in this model arises as follows. In the observable sector
the gauge group remains as in (\ref{ggfsu5model}). In the hidden
sector the NS sector produces the gauge bosons of the
\beq
SU(4)\times U(1)\times SO(4)\times SO(4)
\eeq
subgroup, and the sector $\xi_2=\bone+b_1+b_2+b_3$
produces the vector bosons that
complete the hidden gauge group to $SO(10)\times SO(6)$.
Thus, the four dimensional gauge group is identical
in the two models. The sectors $b_{1,2,3}$, $b_2+b_5$,
$b_1+b_4$ and $S+b_1+b_2+b_4+b_5$ are not affected
by this map, and therefore trivially produce the same
spectrum. The three hidden $SO(16)$ vectorial representations
are now obtained from the sectors $b_{1,2,3}+(2\gamma\oplus\xi_2)$,
and are decomposed under the unbroken hidden $SO(10)\times SO(6)$
gauge group. Thus the spectrum from these sectors is identical
to the one found in the model of [\ref{fsu51}].
Finally, the exotic fractionally charged states
are obtained in this model from the sectors
$      b_3        \pm\gamma\oplus\xi_2$;
$S+b_2+b_3+b_5    \pm\gamma\oplus\xi_2$;
$S+b_2+b_3+b_4+b_5\pm\gamma\oplus\xi_2$;
$S+b_1+b_3+b_4    \pm\gamma\oplus\xi_2$;
$S+b_1+b_3+b_4+b_5\pm\gamma\oplus\xi_2$;
$S+b_1+b_2+b_3+b_5\pm\gamma\oplus\xi_2$,
and, as in the previous model, transform as $4+{\bar 4}$ of the hidden
$SO(6)$ gauge group. Hence, we see that the entire spectrum of the two
models is identical,
with the substitutions
\beqn
 2\gamma^\prime && \rightarrow (2\gamma\oplus\xi_2)~,\cr
 \gamma^\prime  && \rightarrow (\gamma\oplus\xi_2)~,
\eeqn
in sectors that preserve, and break, the observable $SO(10)$
symmetry, respectively.

It ought to be remarked, however, that the map
$\gamma\rightarrow\gamma^\prime$ does not always exist
in the case of the three generation standard--like models. 
The reason being that there are such cases in which the 
modular invariant constraints are not preserved by the map.
Such example are provided by the models of refs \cite{eu,top}.
In these models the assignment in the basis vectors $\{\alpha,\beta\}$,
and $\gamma$, is such that the product $\alpha\cdot\gamma$
among the world--sheet fermions that produce the observable $E_8$
gauge group is 3/2. This means that product between these basis
vectors in the hidden sector has to be half--integral as well.
Thus, as the map $\gamma^\prime\rightarrow\gamma$ removes an even number of
half--integral boundary conditions, it cannot preserve the modular
invariance constraints. Nevertheless, also in these models,
the Higgs and matter sectors still preserve their $E_6$ origins,
as they originate from sectors that preserve the $SO(10)$ symmetry.
Similarly, the models of refs. \cite{lrsmodels,su421} do not
originate from an $N=4$ $SO(12)\times E_8\times E_8$ vacuum,
but rather from $N=4$ $SO(16)\times E_7\times E_7$, and $SO(28)\times E_8$,
respectively. Therefore, in this cases the overlap between $\xi_1$ and 
$\xi_2$ is not empty, and there is no equivalence between the map
and the discreet choice of the phase. However, 
the models of ref. \cite{lrsmodels, su421} do not produce
realistic spectra, as discussed there. The model of ref. \cite{fsu5}
provides an example of a quasi--realistic three generation free fermionic
model, in which the equivalence between the map and the discreet choice
of the phase is applicable. 

\setcounter{footnote}{0}
\section{Conclusions}
I demonstrated in this paper that the utilization of the vector
$2\gamma$ in a large class of
quasi--realistic free fermionic models is equivalent
to setting the GSO projection coefficient between the two spinorial
generators of the observable and hidden $SO(16)$ group factors
$\xi_1$ and $\xi_2$ to
$$c{\xi_1\choose\xi_2}=-1.$$
Although, the equivalence was illustrated in several concrete model,
I conjecture that it is in fact a general equivalence, and arises
from modular properties of the $N=1$ partition function.
Thus, this equivalence applies to the larger class of free fermionic
models. It would be of further interest to examine whether it 
applies on other classes of string compactification,
and what are the precise modular properties that it reflects. 

This results in the projection of the states from the sectors $\xi_1$ and
$\xi_2$ and has important phenomenological implications. At the 
$N=4$ level it results in the breaking of the $E_8\times E_8$
gauge group to $SO(16)\times SO(16)$. In the $N=1$ ten dimensional
level it implies the breaking of $N=1$ supersymmetry. This result
arises in ten dimensions because of the identity $S=\bone+\xi_1+\xi_2$,
where $S$ is the supersymmetry generator. A question of interest in this
respect is whether this phase plays a role in supersymmetry
breaking in lower dimensions. In ref. \cite{fknr} it was argued that free
phases in the partition function may in certain cases be interpreted
as vacuum expectation values of background fields in the effective
field theory description of the string vacuum. A question of interest
from this point of view is whether the GSO phase $c{x_1\choose\xi_2}$
admits such an interpretation.

In the $N=1$ model the choice of the GSO phase (\ref{cxi1xi2minus})
results in the breaking of $E_6\rightarrow SO(10)\times U(1)$.
In this case the 27 multiplet of $E_6$ splits into  spinorial matter
states from the twisted sectors, 
and vectorial matter states from the untwisted sector.
The would vectorial matter states from the twisted sectors
are mapped to vectorial hidden matter states, whereas the untwisted
spinorial states are projected out. In this way, while the $E_6$
symmetry is broken, the models possess an underlying $E_6$ 
grand unifying structure. The mapping of the twisted observable 
vector states into hidden matter states, also results in the projection
of the twisted moduli in these models. An additional consequence
of this breaking is that the $U(1)$ which is embedded in $E_6$
becomes anomalous \cite{anomu1}, which may be an additional indication
for the relevance of this symmetry breaking pattern for supersymmetry
breaking. 
To summarize, understanding the role of the phase $c{xi_1\choose\xi_2}$
may hold the key to understanding some of the key questions
in the relation between string theory and the particle data.

\bigskip
\medskip
\leftline{\large\bf Acknowledgments}
\medskip

I would like to thank the Theoretical Physics Department at Oxford and
the CERN theory group for hospitality while this work was conducted.
This work was supported in part by the PPARC.

\vfill\eject
\appendix
{\textwidth=7.5in
\oddsidemargin=-18mm
\topmargin=-5mm
\renewcommand{\baselinestretch}{1.3}

\section{Model with enhanced symmetries}\label{modelxi1so10spec}

\noindent
\beq
\begin{tabular}{|c|c|rr||c|}\hline
 SEC &   $SU(6) \times
           SU(2)_L\times$   
	                 & $Q_1^\prime$ & $Q_2^\prime$ &  $E_7\times
                                                            SU(2)_{H_1}$  \\
\hline
Neveu --
     & $({15},1)$        &   2   &   2   &   $(1,1)$  \\
 Schwarz    
     & $(\overline{15},1)$
                         &  $-2$   &  $-2$   &   $(1,1)$	\\
  $\oplus$
     & $({15},1)$        &  $-2$   &   2   &   $(1,1)$	\\
  $\xi_1$
     & $(\overline{15},1)$
                         &   2   &  $-2$   &   $(1,1)$	\\
     & $({15},1)$        &   0   &  $-2$   &   $(1,1)$	\\
     & $(\overline{15},1)$
                         &   0   &   2   &   $(1,1)$	\\
     & $({1},1)$         &$\mp2$ &$\pm6$ &   $(1,1)$	\\
     & $({1},1)$         &$\pm2$ &   0   &   $(1,1)$	\\
     & $({1},1)$         &$\pm2$ &$\pm6$ &   $(1,1)$	\\
     & $6\times({1},1)~~~~~~$         
                         &   0   &   0   &   $(1,1)$	\\
\hline
${b}_1\oplus\xi_1$ 
     & $({15},1)$       &   1	 &   1   &   $(1,1)$    \\
     & $({6},2)$        &   1    &   1   &   $(1,1)$	\\
     & $({1},1)$        &  $-3$    &  $-3$   &   $(1,1)$	\\
     & $4\times({1},1)~~~~~~$        
                        &$\pm1$  &$\mp3$ &   $(1,1)$	\\
\hline
${b}_2\oplus\xi_1$ 
     & $({15},1)$       &  $-1$	 &   1   &   $(1,1)$    \\
     & $({6},2)$        &  $-1$  &   1   &   $(1,1)$	\\
     & $({1},1)$        &   3    &  $-3$ &   $(1,1)$	\\
     & $4\times({1},1)~~~~~~$        
                        &$\pm1$  &$\pm3$ &   $(1,1)$	\\
\hline
${b}_3\oplus\xi_1$ 
     & $({15},1)$       &   0	&  $-1$   &    $(1,1)$  \\
     & $({6},2)$        &   0   &  $-1$   &    $(1,1)$   \\
     & $({1},1)$        &   0   &   6   &    $(1,1)$	\\
     & $4\times({1},1)~~~~~~$        
                        &   0   &$\pm1$ &   $(1,1)$	\\
\hline
${b}_4\oplus\xi_1$ 
     & $({15},1)$        &   1	 &   1   &   $(1,1)$  \\
     & $(\bar{6},2)$     &  $-1$   &  $-1$   &   $(1,1)$	\\
     & $({1},1)$         &  $-3$   &  $-3$   &    $(1,1)$	\\
     & $4\times({1},1)~~~~~~$        
                        &$\pm1$  &$\mp3$ &   $(1,1)$	\\
\hline
${b}_5\oplus\xi_1$ 
     & $({15},1)$      &  $-1$     &   1   &   $(1,1)$  \\
     & $(\bar{6},2)$   &   1     &  $-1$   &   $(1,1)$	\\
     & $({1},1)$       &   0     &  $-3$   &    $(1,1)$	\\
     & $4\times({1},1)~~~~~~$        
                        &$\pm1$  &$\mp3$ &   $(1,1)$	\\
\hline
$S+$ 
     & $({15},1)$      &   0	 &  $-1$   &   $(1,1)$  \\
${b}_4 + {b}_5\oplus$ 
     & $(\bar{6},2)$   &   0	 &   1   &   $(1,1)$  \\
$\xi_1$ 
     & $({1},1)$   &   0	 &   6   &   $(1,1)$  \\
     & $4\times({1},1)~~~~~~$   
                     &  $\pm1$	 &   0   &   $(1,1)$  \\
\hline

\end{tabular}
\label{xi1xi2plus}
\eeq
\noindent 
$SO(10)$ preserving spectrum in the model of table [\ref{toyso64model}],
with $c{\xi_1\choose\xi_2}=+1$.
The symbol $\oplus$ is used to denote that the states arise from the
two sectors $a$ \& $a+\xi_1$. Here $SO(10)$ preserving means that these
states arise from sectors that do not contain the basis vector $\alpha$.

\vfill\eject


\noindent
\beq
\begin{tabular}{|c|c|rr||c|}\hline
 SEC &   $SU(6) \times
           SU(2)_L$   
	                 & $Q_1^\prime$ & $Q_2^\prime$ &  $E_7\times
                                                            SU(2)_{H_1}$  \\
\hline
$S+b_2+b_4+$
     & $({6},1)$       &   0	 &  $-2$   &   $(1,2)$  \\
$\alpha \oplus\xi_1$ 
     & $({1},2)$       &  $-2$	 &   0   &   $(1,2)$  \\
\hline
$b_1+b_4+b_5+$
     & $({6},1)$       &  $-1$	 &   1   &   $(1,2)$  \\
$\alpha \oplus\xi_1$ 
     & $({1},2)$       &   1	 &   3   &   $(1,2)$  \\
\hline
$b_1+b_2+b_4+$
     & $({6},1)$       &  $-1$	 &   1   &   $(1,2)$  \\
$\alpha \oplus\xi_1$ 
     & $({1},2)$       &  $-1$	 &  $-3$   &   $(1,2)$  \\
\hline
$b_2+b_3+b_5+$
     & $({6},1)$       &   0	 &  $-1$   &   $(1,2)$  \\
$\alpha \oplus\xi_1$ 
     & $({1},2)$       &   1	 &   0   &   $(1,2)$  \\
\hline
$S+b_3+b_5+$
     & $({6},1)$       &   1	 &   1   &   $(1,2)$  \\
$\alpha \oplus\xi_1$ 
     & $({1},2)$       &   1	 &  $-3$   &   $(1,2)$  \\
\hline
$S+b_1+b_2+b_3+b_4+$
     & $({6},1)$       &   1	 &   1   &   $(1,2)$  \\
$\alpha \oplus\xi_1$ 
     & $({1},2)$       &  $-1$	 &   3   &   $(1,2)$  \\
\hline
$b_2+b_3+b_5+$
     & $({6},1)$       &   0	 &  $-1$   &   $(1,2)$  \\
$\alpha \oplus\xi_1$ 
     & $({1},2)$       &   1	 &   0   &   $(1,2)$  \\
\hline
$S+b_3+b_5+$
     & $({6},1)$       &   1	 &   1   &   $(1,2)$  \\
$\alpha \oplus\xi_1$ 
     & $({1},2)$       &   1	 &  $-3$   &   $(1,2)$  \\
\hline

\end{tabular}
\label{xi1xi2pluse}
\eeq
\noindent
$SO(10)$ breaking spectrum in the model of table [\ref{toyso64model}],
with $c{\xi_1\choose\xi_2}=+1$.

\vfill\eject

\section{Model with $c{\xi_1\choose\xi_2}=-1$}\label{modelxi2so10spec}

\noindent
\beq
\begin{tabular}{|c|c|rrr||c|}\hline
 SEC &   $SU(4) \times$ & $Q^\prime_6$ & 
	                   $Q^\prime_1$ & 
			   $Q^\prime_2$ &              $SO(12)\times$
                                                         \\
      &  $SU(2)_L\times 
            SU(2)_R$    &    &    &     &     $SU(2)_{H_1}\times
                                                      SU(2)_{H_2}$\\
\hline
Neveu --
     & $({6},1,1)$       &$\pm2$ &$\pm2$ &$\pm2$ &   $(1,1,1)$  \\
 Schwarz    
     & $({6},1,1)$       &$\pm2$ &$\mp2$ &$\pm2$ &   $(1,1,1)$	\\
     & $({6},1,1)$       &$\mp2$ &   0   &$\pm4$ &   $(1,1,1)$	\\
     & $({1},1,1)$       &   0   &$\pm2$ &$\pm6$ &   $(1,1,1)$	\\
     & $({1},1,1)$       &$\pm4$ &$\mp2$ &$\mp2$ &   $(1,1,1)$	\\
     & $({1},1,1)$       &$\pm4$ &   0   &$\mp4$ &   $(1,1,1)$	\\
     & $({1},1,1)$       &   0   &$\pm4$ &  0    &   $(1,1,1)$	\\
     & $({1},1,1)$       &   0   &$\pm2$ &$\pm6$ &   $(1,1,1)$	\\
     & $({1},1,1)$       &$\pm4$ &$\pm2$ &$\mp2$ &   $(1,1,1)$	\\
     & $6\times({1},1)~~~~~~$         
                         &   0   &   0   &   $(1,1)$	\\
\hline
${b}_1$ 
     & $({4},2,1)$       &   1	 &   1   &  1    &   $(1,1,1)$  \\
     & $(\bar{4},1,2)$   &   1   &   1   &  1    &   $(1,1,1)$	\\
\hline
${b}_1+\xi_1$ 
     & $({1},1,1)$       &   0	 &  $-1$   &  3    &   $(12,1,1)$  \\
     & $({1},1,1)$       &   0   &  $-1$   &  3    &   $( 1,2,2)$	\\
\hline
${b}_2$ 
     & $(    {4},2,1)$   &   1	 &  $-1$   &  1    &   $(1,1,1)$  \\
     & $(\bar{4},1,2)$   &   1   &  $-1$   &  1    &   $(1,1,1)$	\\
\hline
${b}_2+\xi_1$ 
     & $({1},1,1)$       &   0	 &   1   &  3    &   $(12,1,1)$  \\
     & $({1},1,1)$       &   0   &   1   &  3    &   $( 1,2,2)$	\\
\hline
${b}_3$ 
     & $(    {4},2,1)$   &  $-1$	 &   0   &  2    &   $(1,1,1)$  \\
     & $(\bar{4},1,2)$   &  $-1$   &   0   &  2    &   $(1,1,1)$	\\
\hline
${b}_3+\xi_1$ 
     & $({1},1,1)$       &   1	 &   0   &  1    &   $(12,1,1)$  \\
     & $({1},1,1)$       &   1   &   0   &  1    &   $( 1,2,2)$	\\
\hline
${b}_4$ 
     & $(\bar{4},2,1)$   &  $-1$	 &  $-1$   & $-1$    &   $(1,1,1)$  \\
     & $(\bar{4},1,2)$   &   1   &   1   &  1    &   $(1,1,1)$	\\
\hline
${b}_4+\xi_1$ 
     & $({1},1,1)$       &   0	 &  $-1$   &  3    &   $(12,1,1)$  \\
     & $({1},1,1)$       &   0   &   1   & $-3$    &   $( 1,2,2)$	\\
\hline
${b}_5$ 
     & $(\bar{4},2,1)$   &  $-1$	 &   1   & $-1$    &   $(1,1,1)$  \\
     & $(\bar{4},1,2)$   &   1   &  $-1$   &  1    &   $(1,1,1)$	\\
\hline
${b}_5+\xi_1$ 
     & $({1},1,1)$       &   0	 &  $-1$   & $-3$&   $(12,1,1)$  \\
     & $({1},1,1)$       &   0   &   1   &  3    &   $( 1,2,2)$	\\
\hline
$S+{b}_4 + {b}_5$ 
     & $({6},1,1)$   &  $-2$	 &   0   & $-2$  &   $(1,1,1)$  \\
     & $({1},2,2)$   &   2	 &   0   &  2    &   $(1,1,1)$  \\
     & $({1},1,1)$   &   4	 &   0   & $-2$  &   $(1,1,1)$  \\
     & $({1},1,1)$   &   0	 &   0   &  6    &   $(1,1,1)$  \\
     & $4 \times ~({1},1,1)~~~~~~$   
                     &   0    	 & $\pm2$&  0    &   $(1,1,1)$  \\
\hline
\end{tabular}
\label{xi1xi2minus}
\eeq
\noindent
$SO(10)$ preserving spectrum in the model of table [\ref{toyso64model}],
with $c{\xi_1\choose\xi_2}=-1$.

\vfill\eject


\beq
\begin{tabular}{|c|c|rrr||c|}\hline
 SEC &   $SU(4) \times$ & $Q^\prime_6$ & 
	                   $Q^\prime_1$ & 
			   $Q^\prime_2$ &              $SO(12)\times$
                                                         \\
      &  $SU(2)_L\times 
            SU(2)_R$    &    &    &     &     $SU(2)_{H_1}\times
                                                       SU(2)_{H_2}$\\
\hline
${b}_1+b_4+$ 
     & $({4},1,1)$       &   0   &  $-1$   &  3    &   $( 1,2,1)$  \\
$b_5+\alpha+\xi_1$ 
     &                   &       &         &       &               \\
\hline
${b}_1+b_4+ $
     & $({1},2,1)$       &   0	 &   1   &  3    &   $(1,1,2)$  \\
$b_5+\alpha$
     & $({1},1,2)$       &  $-2$   &  $-1$   &  1    &   $(1,1,2)$	\\
\hline
${b}_1+b_2+$ 
     & $({4},1,1)$       &   1   &  $-1$   &  1    &   $( 1,2,1)$  \\
$b_4+\alpha+\xi_1$ 
     &                   &       &         &       &               \\
\hline
${b}_1+b_2+$
     & $({1},2,1)$       &   0	 &  $-1$   & $-3$    &   $(1,1,2)$  \\
$b_4+\alpha$ 
     & $({1},1,2)$       &  $-2$   &  $-1$   &  1    &   $(1,1,2)$  \\
\hline
$S+b_2+$
     & $({4},1,1)$       &   1	 &   0   & $-2$    &   $(1,1,2)$  \\
$b_4+\alpha$ 
     &                   &    	 &       &         &              \\
\hline
$S+b_2+$
     & $({1},2,1)$       &   0	 &  $-1$   &  0    &   $(1,1,2)$  \\
$b_4+\alpha+\xi_1$ 
     & $({1},1,2)$       &  $-2$	 &   0   & $-2$    &   $(1,1,2)$  \\
\hline
${b}_2+b_3+$ 
     & $(    {4},1,1)$   &  $-1$	 &   0   &  2    &   $(1,2,1)$  \\
$b_5+\alpha+\xi_1$ 
     &                   &      	 &       &       &              \\
\hline
${b}_2+b_3+$
     & $(    {1},2,1)$   &   2	 &   0   &  2    &   $(1,1,2)$  \\
$b_5+\alpha$ 
     & $(    {1},1,2)$   &   0   &  $-2$   &  0    &   $(1,1,2)$	\\
\hline
$S+{b}_1+$
     & $(    {4},1,1)$   &  $-1$	 &  $-1$   & $-1$    &   $(1,1,2)$  \\
$b_3+b_4+$
     &                   &       	 &         &         &              \\
$b_5+\alpha+\xi_1$ 
     &                   &       	 &         &         &              \\
\hline
$S+{b}_2+b_3+$
     & $(    {1},2,1)$   &   2	 &  $-1$   & $-1$    &   $(1,2,1)$  \\
$b_4+b_5+\alpha$ 
     & $(    {1},1,2)$   &   0   &   2   & $-3$    &   $(1,2,1)$	\\
\hline
$S+b_3+b_5+\alpha$ 
     & $(\bar{4},1,1)$   &   1	 &   1   &  1    &   $(1,1,2)$  \\
     &                   &    	 &       &       &              \\
\hline
$S+b_3+b_5+$
     & $(    {1},2,1)$   &   2	 &  $-1$   & $-1$    &   $(1,2,1)$  \\
$\alpha+\xi_1$ 
     & $(    {1},1,2)$   &   0   &   2   & $-3$    &   $(1,2,1)$	\\
\hline

\end{tabular}
\label{xi1xi2minuse}
\eeq
$SO(10)$ breaking spectrum in the model of table [\ref{toyso64model}],
with $c{\xi_1\choose\xi_2}=-1$.

\vfill\eject

\section{Model with 2$\gamma$}\label{model2gso10spec}

\noindent
\beq
\begin{tabular}{|c|c|rrr||c|}\hline
  SEC &   $SU(4) \times$ & $Q^\prime_6$ & 
	                   $Q^\prime_1$ & 
			   $Q^\prime_2$ &              $SO(12)\times$
                                                         \\
      &  $SU(2)_L\times 
            SU(2)_R$    &    &    &     &     $SU(2)_{H_1}\times
                                                       SU(2)_{H_2}$\\
\hline
Neveu --
     & $({6},1,1)$       & $\pm2$&$\pm2$ &$\pm2$ &   $(1,1,1)$  \\
 Schwarz    
     & $({6},1,1)$       &$\pm2$ &$\mp2$ &$\pm2$ &   $(1,1,1)$	\\
     & $({6},1,1)$       &$\mp2$ &   0   &$\pm4$ &   $(1,1,1)$	\\
     & $({1},1,1)$       &   0   &$\pm2$ &$\pm6$ &   $(1,1,1)$	\\
     & $({1},1,1)$       &$\pm4$ &$\mp2$ &$\mp2$ &   $(1,1,1)$	\\
     & $({1},1,1)$       &$\pm4$ &   0   &$\mp4$ &   $(1,1,1)$	\\
     & $({1},1,1)$       &   0   &$\pm4$ &  0    &   $(1,1,1)$	\\
     & $({1},1,1)$       &   0   & $\pm2$&$\pm6$ &   $(1,1,1)$	\\
     & $({1},1,1)$       &$\pm4$ &$\pm2$ &$\mp2$ &   $(1,1,1)$	\\
     & $6\times({1},1)~~~~~~$         
                         &   0   &   0   &   $(1,1)$	\\
\hline
${b}_1$ 
     & $({4},2,1)$       &   1	 &   1   &  1    &   $(1,1,1)$  \\
     & $(\bar{4},1,2)$   &   1   &   1   &  1    &   $(1,1,1)$	\\
\hline
${b}_1+(2\gamma\oplus\xi_2)$ 
     & $({1},1,1)$   &   0	 &  $-1$ &  3    &   $(12,1,1)$  \\
     & $({1},1,1)$   &   0	 &  $-1$ &  3    &   $(1,2,2)$  \\
\hline
${b}_2$ 
     & $({4},2,1)$       &   1	 & $-1$  &  1    &   $(1,1,1)$  \\
     & $(\bar{4},1,2)$   &   1   & $-1$  &  1    &   $(1,1,1)$	\\
\hline
${b}_2+(2\gamma\oplus\xi_2)$ 
     & $({1},1,1)$   &   0	 &   1   &  3    &   $(12,1,1)$  \\
     & $({1},1,1)$   &   0	 &   1   &  3    &   $(1,2,2)$  \\
\hline
${b}_3$ 
     & $({4},2,1)$       & $-1$	 &   0   &  2    &   $(1,1,1)$  \\
     & $(\bar{4},1,2)$   & $-1$  &   0   &  2    &   $(1,1,1)$	\\
\hline
${b}_3+(2\gamma\oplus\xi_2)$ 
     & $({1},1,1)$   &   1	 &   0   &  1    &   $(12,1,1)$  \\
     & $({1},1,1)$   &   1	 &   0   &  1    &   $(1,2,2)$  \\
\hline
${b}_4$ 
     & $(\bar{4},2,1)$   &  $-1$ & $-1$  & $-1$  &   $(1,1,1)$  \\
     & $(\bar{4},1,2)$   &   1   &   1   &  1    &   $(1,1,1)$	\\
\hline
${b}_4+(2\gamma\oplus\xi_2)$ 
     & $({1},1,1)$   &   0	 & $-1$    &  3      &   $(12,1,1)$  \\
     & $({1},1,1)$   &   0	 &   1     &  3      &   $(1,2,2)$  \\
\hline
${b}_5$ 
     & $(\bar{4},2,1)$   &  $-1$ &   1   &$-1$   &   $(1,1,1)$  \\
     & $(\bar{4},1,2)$   &   1   &$-1$   &  1    &   $(1,1,1)$	\\
\hline
${b}_5+(2\gamma\oplus\xi_2)$ 
     & $({1},1,1)$   &   0	 &$-1$   & $-3$  &   $(12,1,1)$  \\
     & $({1},1,1)$   &   1	 &   1   &  3    &   $(1,2,2)$  \\
\hline
$S+{b}_4 + {b}_5$ 
     & $({6},1,1)$   &  $-2$	 &   0   &$-2$   &   $(1,1,1)$  \\
     & $({1},2,2)$   &   2	 &   0   &  2    &   $(1,1,1)$  \\
     & $({1},1,1)$   &   4	 &   0   & $-2$  &   $(1,1,1)$  \\
     & $({1},1,1)$   &   0	 &   0   &  6    &   $(1,1,1)$  \\
     & $4 \times ~({1},1,1)~~~~~~$   
                     &   0	 & $\pm2$&  0    &   $(1,1,1)$  \\
\hline

\end{tabular}
\label{xi2gamma}
\eeq
\noindent
$SO(10)$ preserving spectrum in the model of table [\ref{toyso64model}],
with the substitution $\xi_1\rightarrow2\gamma$.

\vfill\eject


\noindent
\beq
\begin{tabular}{|c|c|rrr||c|}\hline
 SEC &   $SU(4) \times$ & $Q^\prime_6$ & 
	                   $Q^\prime_1$ & 
			   $Q^\prime_2$ &              $SO(12)\times$
                                                         \\
      &  $SU(2)_L\times 
            SU(2)_R$    &    &    &     &     $SU(2)_{H_1}\times
                                                       SU(2)_{H_2}$\\
\hline
${b}_1+b_4+b_5+$ 
     & $({4},1,1)$       &   1   &  $-1$   &  1    &   $( 1,2,1)$  \\
$\alpha+(\xi_2+2\gamma)$ 
     &                   &       &         &       &               \\
\hline
${b}_1+b_4+$
     & $({1},2,1)$       &   0	 &   1   &  3    &   $(1,1,2)$  \\
$b_5+\alpha$ 
     & $({1},1,2)$       &  $-2$   &  $-1$   &  1    &   $(1,1,2)$	\\
\hline
${b}_1+b_2+b_4+$ 
     & $({4},1,1)$       &   1   &  $-1$   &  1    &   $( 1,2,1)$  \\
$\alpha+(\xi_2+2\gamma)$ 
     &                   &       &         &       &               \\
\hline
${b}_1+b_2+$
     & $({1},2,1)$       &   0	 &  $-1$   & $-3$    &   $(1,1,2)$  \\
$b_4+\alpha$ 
     & $({1},1,2)$       &  $-2$   &  $-1$   &  1    &   $(1,1,2)$	\\
\hline
$S+b_2+$ 
     & $({4},1,1)$       &   1	 &   0   & $-2$    &   $(1,1,2)$  \\
$b_4+\alpha$ 
     &                   &    	 &       &         &              \\
\hline
$S+b_2+b_4+$
     & $({1},2,1)$       &   0	 &  $-1$   &  0    &   $(1,1,2)$  \\
$\alpha+(\xi_2+2\gamma)$ 
     & $({1},1,2)$       &  $-2$	 &   0   & $-2$    &   $(1,1,2)$  \\
\hline
${b}_2+b_3+b_5+$ 
     & $(    {4},1,1)$   &  $-1$	 &   0   &  2    &   $(1,2,1)$  \\
$\alpha+(\xi_2+2\gamma)$ 
     &                   &       	 &       &       &              \\
\hline
${b}_2+b_3+$
     & $(    {1},2,1)$   &   2	 &   0   &  2    &   $(1,1,2)$  \\
$b_5+\alpha$ 
     & $(    {1},1,2)$   &   0   &  $-2$   &  0    &   $(1,1,2)$	\\
\hline
$S+{b}_1+b_3+$ 
     & $(    {4},1,1)$   &  $-1$	 &  $-1$   & $-1$    &   $(1,1,2)$  \\
$b_4+b_5+\alpha+$ 
     &                   &      	 &         &         &              \\
$(\xi_2+2\gamma)$ 
     &                   &      	 &         &         &              \\
\hline
$S+{b}_2+b_3+$
     & $(    {1},2,1)$   &   2	 &  $-1$   & $-1$    &   $(1,2,1)$  \\
$b_4+b_5+\alpha$ 
     & $(    {1},1,2)$   &   0   &   2   & $-3$    &   $(1,2,1)$	\\
\hline
$S+b_3+$ 
     & $(\bar{4},1,1)$   &   1	 &   1   &  1    &   $(1,1,2)$  \\
$b_5+\alpha$ 
     &                   &    	 &       &       &              \\
\hline
$S+b_3+b_5+                      $ 
     & $(    {1},2,1)$   &   2	 &  $-1$   & $-1$    &   $(1,2,1)$  \\
$\alpha+(\xi_2+2\gamma)$ 
     & $(    {1},1,2)$   &   0   &   2   & $-3$    &   $(1,2,1)$	\\
\hline

\end{tabular}
\label{xi2gammae}
\eeq
\noindent
$SO(10)$ breaking spectrum in the model of table [\ref{toyso64model}],
with the substitution $\xi_1\rightarrow2\gamma$.

\hfill\vfill\eject}

\vfill\eject

\bigskip
\medskip

\bibliographystyle{unsrt}

\vfill\eject
\end{document}